\documentclass[12pt]{article}
\usepackage{amsmath}
\usepackage{amssymb}
\usepackage{amsfonts}
\usepackage{amscd}
\usepackage{delarray}
\usepackage{graphicx}

\def\bd{\begin{displaystyle}}
\def\ed{\end{displaystyle}}
\def\ba{\begin{eqnarray}}
\def\ea{\end{eqnarray}}
\def\bea{\begin{eqnarray*}}
\def\eea{\end{eqnarray*}}

\let\BBox\Box
\def\Box{$\BBox$}

\begin{document}

\title{Computation of entropy increase for Lorentz gas and hard disks}
\author{M. Courbage and  S. M. Saberi Fathi }
\date{\today}

\maketitle \centerline{\it  Laboratoire Mati\`ere et Syst\`emes
Complexes (MSC) } \centerline{\it UMR 7057 CNRS et Universit\'e
Paris 7- Denis Diderot } \centerline{\it Case 7020, Tour
24-14.5\`eme \'etage, 4, Place Jussieu} \centerline{\it 75251 Paris
Cedex 05 / FRANCE} \centerline{\it emails : courbage@ccr.jussieu.fr,
saberi@ccr.jussieu.fr} \vspace{1cm}

{\bf\footnotesize Abstract. }{\footnotesize   Entropy functionals
are computed for non-stationary distributions of particles of
Lorentz gas and hard disks. The distributions consisting of beams
of particles are found to have the  largest amount of entropy and
entropy increase. The computations show exponentially monotonic
increase during initial time of rapid approach to equilibrium. The
rate of entropy increase is bounded by sums of positive Lyapounov
exponents.}

\section{Introduction}
\indent

\begin{figure}[!h]
\begin{center}
  \centering
  \includegraphics[width=100mm]{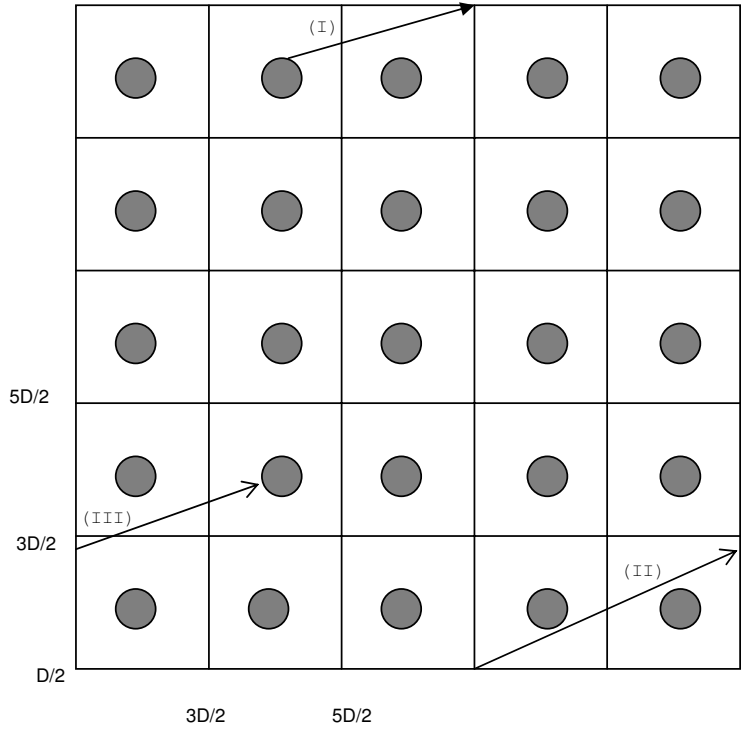}\\
 \caption{\footnotesize{The motion of the particle on a toric billiard.}}\label{collisionfb}
 \end{center}
\end{figure}

 The H-theorem for dynamical systems describes the approach
to equilibrium, the irreversibility and entropy increase for
deterministic evolutions. Suppose that a dynamical transformation
$T$ on a phase space $X$ has some "equilibrium" measure $\mu$,
invariant under $T$, i.e. $\mu(T^{-1}E)=\mu(E)$ for all measurable
subsets $E$ of $X$. Suppose also that there is some mixing type
mechanism of the approach to equilibrium for $T$, i.e. there is a
sufficiently large family of non-equilibrium measures $\nu$ such
that
$\nu_{t}(E)=:\nu(T^{-t}E)\rightarrow_{t\rightarrow\infty}\mu(E)$
for all E. Then, the H-theorem means the existence of a negative
entropy functional $S(\nu_{t})$ which increases monotonically with
$t$ to zero, being attained only for $\nu=\mu$. The existence of
such functional in measure-theoretical dynamical systems has been
the object of several investigations during last decades  see
\cite{C}- \cite{CP}, \cite{gar}, \cite{gold, MPC, sin}). Here we study
this problem for the Lorentz gas and hard disks. The dynamical and
stochastic properties of the Lorentz gas in two dimensions which we
consider here was investigated by Sina\"{i} and Bunimovich as an ergodic
dynamical system
\cite{sin1,BCS, CYo}. Other transport properties have been also studied
numerically (see
\cite{gasp},
\cite{zas}).  This  is a system of non interacting particles
moving with constant velocity and being elastically reflected from
periodically distributed scatterers. The scatterers are fixed
disks. On account of the absence of interactions between particles
the system is reduced to the motion of one billiard ball. We shall
investigate the entropy increase under the effect of collisions of
the particles with the obstacles.  For this purpose, we consider
the map $T$ which associates to an ingoing state of a colliding
particle the next ingoing colliding state. The particle  moves on
an infinite plane, periodically divided into squares of side $D$
called "cells", on the center of which are fixed the scatterers of
radius $a$ ( Fig.~\ref{collisionfb}). The ingoing colliding state
is described by an ingoing unitary velocity arrow  at  some point
of the disk. To a colliding arrow $\textbf{V}_{1}(P_1)$ at point
$P_1$ on the boundary of the disk the map associates the next
colliding arrow $\textbf{V}_2(P_2)$ according to elastic
reflection law. Thus, the collision map does not take into account
the free evolution
between  successive collisions. \\
 Let $\nu$ be a non-equilibrium measure, which means
that $\nu$ is a non invariant measure approaching the equilibrium
$\mu$ in the future. It is mathematically possible to define a
non-equilibrium entropy for a family of such measures, using
conditional expectations (i.e. a generalized averaging)  relatively to the
some remarkable partitions, namely the contracting fibers of the
hyperbolic dynamics \cite{C}.
\indent However, in our numerical simulations some given finite
precision is needed, so that we consider partitions  into cells
with positive $\mu$-measure. Here, we use slightly similar entropy
functionals. Starting from the
non-equilibrium initial distribution $\nu$, and denoting by
$\mathcal{P}$ such partition formed by cells
($\mathcal{P}_1$,$\mathcal{P}_2$, ..., $\mathcal{P}_n$) and by
$\nu_t(\mathcal{P}_i)=\nu\circ T^{-t}(\mathcal{P}_i)$,  the
probability at time $t$ for the system to be in the cell
$\mathcal{P}_i$ and  such that
$\nu(\mathcal{P}_i)\neq\mu(\mathcal{P}_i)$ for some $i$,  the
approach to equilibrium implies that
$\nu_t(\mathcal{P}_i)\rightarrow\mu(\mathcal{P}_i)$ as
$t\rightarrow\infty$ for any $i$. The entropy functional will be
defined by:
\begin{equation}
\mathcal{S}(t,\nu,\mathcal{P})=-\sum_{i=1}^{N}\nu_t(\mathcal{P}_i)
\ln(\frac{\nu_t(\mathcal{P}_i)}{\mu(\mathcal{P}_i)}) := -
\mathcal{H}(t,\nu,\mathcal{P})\label{eq1}
\end{equation}
which we simply denote here after $\mathcal{S}(t)$. The
H-functional (\ref{eq1}) is maximal when the initial distribution
is concentrated on only one cell and minimal if and only if
$\nu_t(\mathcal{P}_i)=\mu(\mathcal{P}_i),\forall i $. These
properties are shown straightforwardly.\\
 This formula describes the relative entropy of the
non-equilibrium measure $\nu_t$ with respect to $\mu$ for the
observation associated to $\mathcal{P}$. It coincides with  the
information theoretical concept of relative entropy of a probability
vector $(p_i)$ with respect to another probability vector $(q_i)$
defined as follows: $- \ln p_i$ being the information of the
$i^\mathrm{th}$ issue under the first distribution, $-\sum_i p_i
\ln(\frac{p_i}{q_i})$, is equal to the average uncertainty gain of
the experience  $(p_i)$ relatively to $(q_i)$.\\
\indent A condition under which formula (\ref{eq1})  shows a
monotonic increase with respect to $t$ is that the process
$\nu_t(\mathcal{P}_i)=\nu\circ T^{-t}(\mathcal{P}_i)$ verifies the
Chapman-Kolmogorov equation valid for Markov chains and other infinite
memory chains. For a dynamical system, this condition is hardly verified
for given partition
$\mathcal{P}$. However, the very rapid mixing leads to a monotonic
increase of the above entropy,  at least during some initial stage, which
can be compared with the relaxation stage in gas theory.\\
\indent In this paper, we will first compute the entropy increase
for some remarkable non-equilibrium distributions over the phase
space of the Sina\"{i} billiard. The billiard system is a hyperbolic
system (with many singularity lines) and, in order to have a rapid
mixing,  we will consider initial distributions supported by the
expanding fibers. Such initial measures have been used in
\cite{C,CP, sin}.   For the billiard the expanding fibers are well
approximated by particles with parallel arrows velocity. We call
this class of initial ensemble beams of particles. We first compute
the entropy increase under the collision map for these initial
distributions.   We will consider  finite uniform partitions of the
phase space as explained below.  The entropy functional will be
defined through (\ref{eq1}). For this purpose, the phase space of
the collision map is described using two angles $(\beta, \psi)$,
where $\beta$ is the angle between the outer normal at $P$ and the
incoming arrows $\textbf{V}(P)$, $\beta\in[0,\frac{\pi}{2}[$, and
$\psi \in [0,\pi]$ is the angle between $x$-axis and the outer
normal at $P$. Thus, the collision map induces a map: $(\beta_1,
\psi_1)\rightarrow(\beta_2, \psi_2)$ (see Fig. ~\ref{collisionc})
and we shall first use a uniform partition of the $(\beta, \psi)$
space. The computation shows that whatever is the coarsening of
these partitions the entropy has the monotonic property in the
initial stage. It is clear that, along mixing process, the initial
distribution will spread over all cells almost reaching the
equilibrium value. Physically, this process is directed by the
strong instability, that is expressed
by the positive Lyapounov exponent.\\
\indent We also consider the relation of the rate of increase of
the entropy functionals and Lyapounov exponents of the Lorentz
gas. Our computation shows that this relation is expressed by an
inequality
\begin{equation}
\max(\mathcal{S}(n+1)-\mathcal{S}(n))\equiv\bigtriangleup\mathcal{S}
\leq\sum_{\lambda_i\geq0}\lambda_i
\end{equation}
where the "$\max$" is taken over $n$, which means that the K-S
entropy is an upper bound of the rate of increase of this functional.\\
\indent In section 3, we shall  consider another phase space and
another partitions associated to spatial extension of the motion
of the Lorentz gas. Here the space in which moves a particle is a
large torus divided into rectangular cells, in the center of each
cell there is one disk. Denoting the total number of cells by $n$
and the number of  particles initially distributed in only one
region, by $N$, and following them until each  executes $ t$
collisions with obstacles, we compute the probability that a
particle is located in the $i\mathrm{th}$ cell as given by:
$$\rho_i(t)=\frac{\mathrm{Number \,of \, particles\, in \,cell \,\emph{i}\, having \,
made \,\emph{t}\,collisions} }{N}$$ The equi-distribution of the
cells leads to take, as equilibrium measure, $\mu_i=\frac{1}{n}$,
so that this "space entropy" is defined by:
\begin{eqnarray}
\mathcal{S}_{sp}(t)=-\sum_{i=1}^{n}\rho_i(t) \ln (\rho_i(t)
n)\label{eq2}
\end{eqnarray}
The maximum of absolute value of this entropy is equal to $-\ln n$.
So we normalize as follows:
\begin{eqnarray}
{\bf s}_{sp}(t) =\frac{\mathcal{S}_{sp}(t)}{\ln n}\label{eq2-1}
\end{eqnarray}

In section \ref{sechs} we shall consider the hard disks systems.
We shall compute an entropy functional similar to the
space-entropy on extended torus with several cells. The
probabilities are defined as for the space entropy in the Lorentz
gas. We shall also do some comparisons of the $H$-theorem with the
sum of normalized positive Lyapounov exponents.
\section{Entropy for collision map}
\begin{figure}[!h]
\begin{center}
  \centering
  \includegraphics[width=65mm]{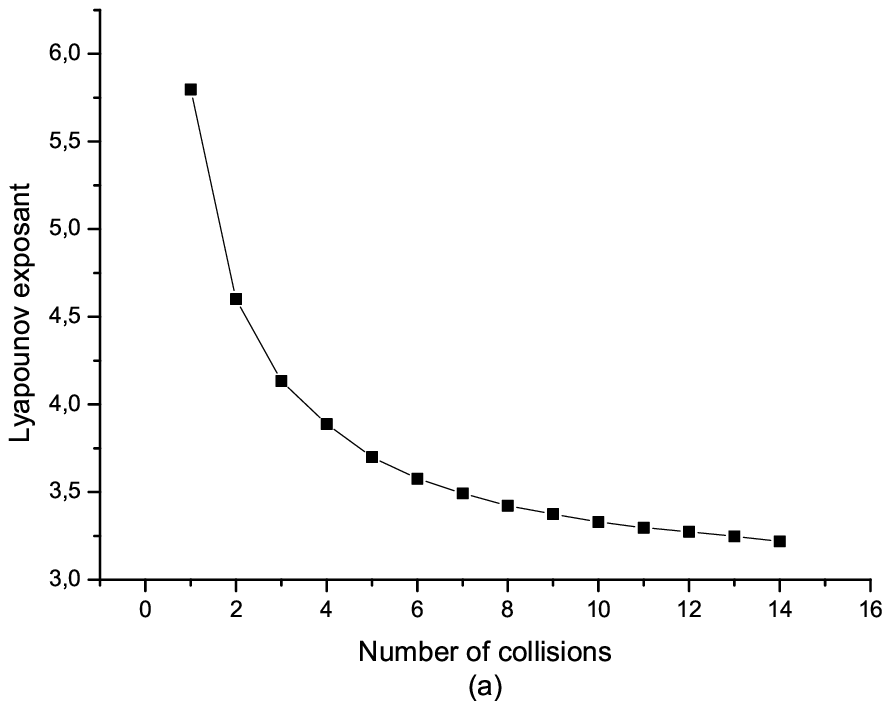}
  \includegraphics[width=65mm]{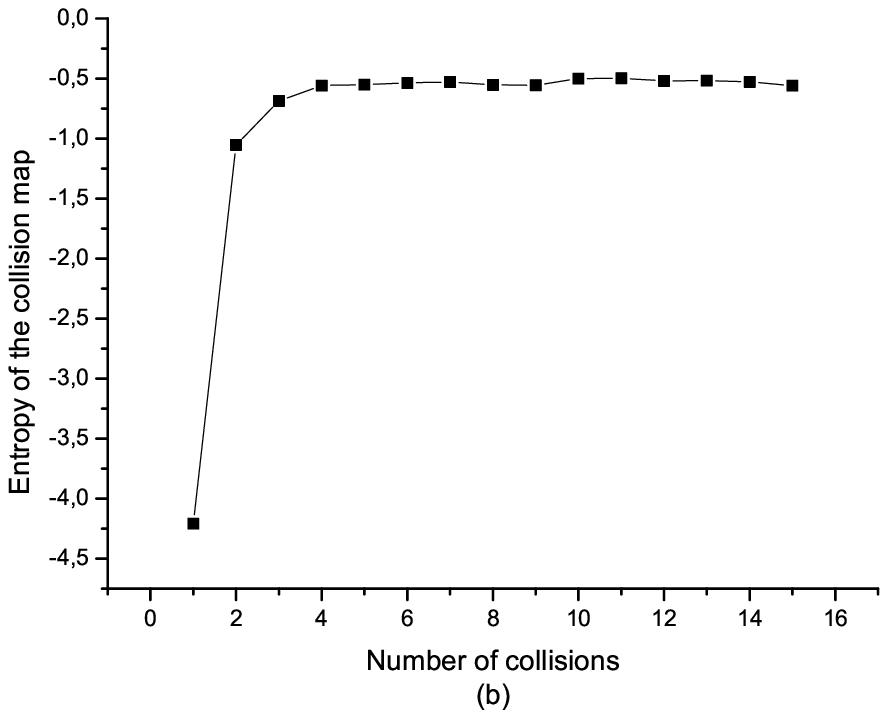}
\caption{\footnotesize{(a) Lyapounov exponent and (b) entropy of
the collision map, versus of number of collisions for each
particle. We see that the maximum of the entropy increase between
two collisions is less than of the value of the Lyapounov
exponent.}}\label{lyem}
 \end{center}
\end{figure}

\begin{figure}[!h]
\begin{center}
  \centering
  \includegraphics[width=65mm]{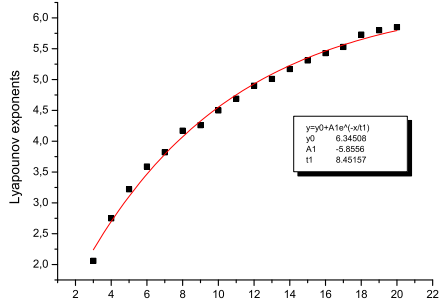}
\caption{\footnotesize{Lyapounov exponent versus
$D/a$.}}\label{lyada}
 \end{center}
\end{figure}

\begin{figure}[!h]
\begin{center}
  \centering
  \includegraphics[width=40mm]{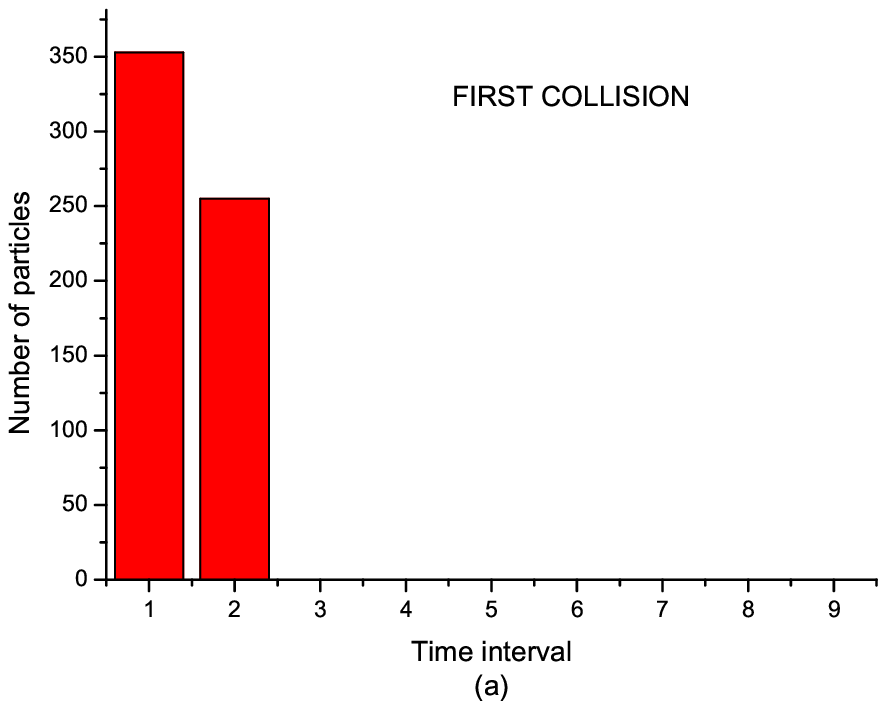}
  \includegraphics[width=40mm]{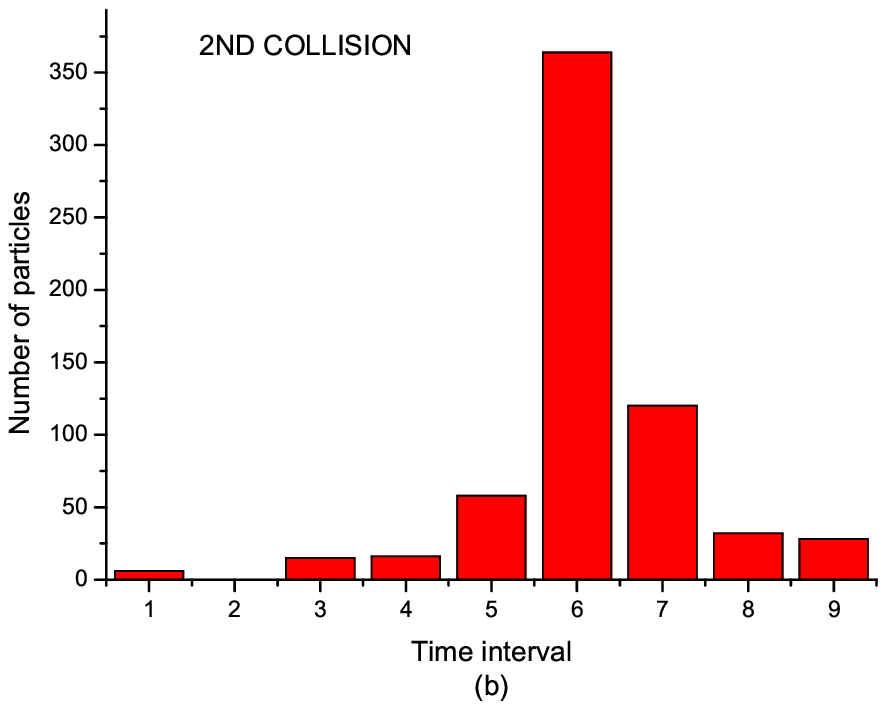}
  \includegraphics[width=40mm]{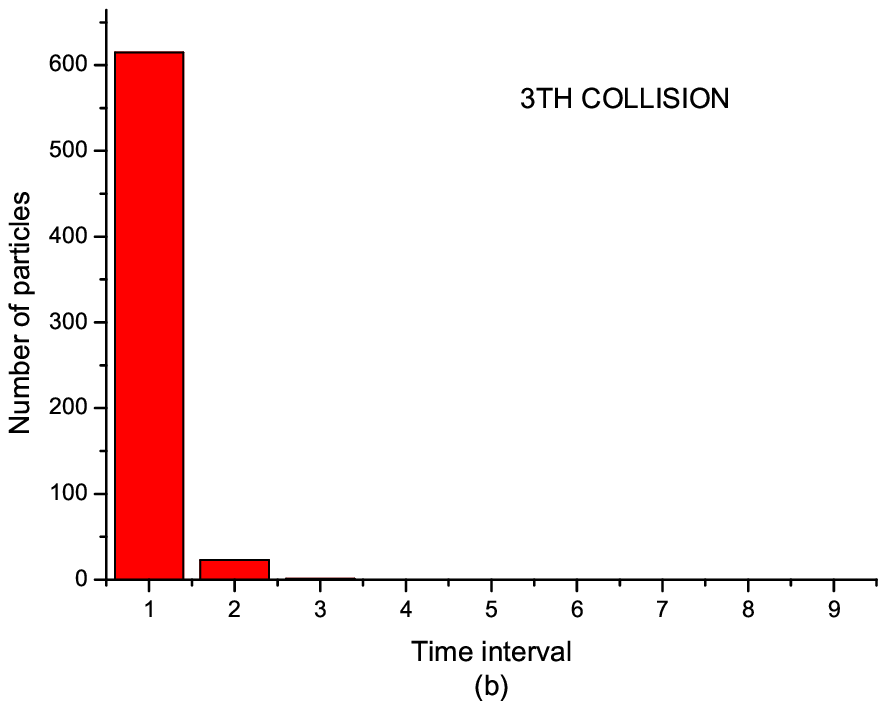}
\caption{\footnotesize{Free time histogram for (a) first, (b)
second and (c) third collision.}}\label{timeh}
 \end{center}
\end{figure}

\indent The entropy for the collision map is computed for a beam
of $N$ particles on a toric checkerboard with $n$ cells. We start
to calculate the entropy, just after all particles have executed
the first collision. In this computation, all particles have the
same initial velocity and are distributed in a small part of one
cell. For each particle we determine the first obstacle and the
angles $(\beta_1, \psi_1)$ of the velocity incoming vector
$\textbf{V}_1(P_1)$ ( see the figures  given in the appendix). For
a uniform partition $\mathcal{P}$ of the space of the variables
$(\beta, \psi)$, the entropy $\mathcal{S}(t) $is computed
iteratively just after all particles have executed the
$\emph{t}^{\mathrm{th}}$ collision. We use the formula
$(\ref{eq1})$ where
\begin{equation}
\mu(\mathcal{P}_i)=\int_{\beta_i}^{\beta_{i+1}}\int_{\psi_i}^{\psi_{i+1}}\cos\beta
d\beta d\psi
\end{equation}
\begin{figure}[!h]
\begin{center}
  \centering
  \includegraphics[width=65mm]{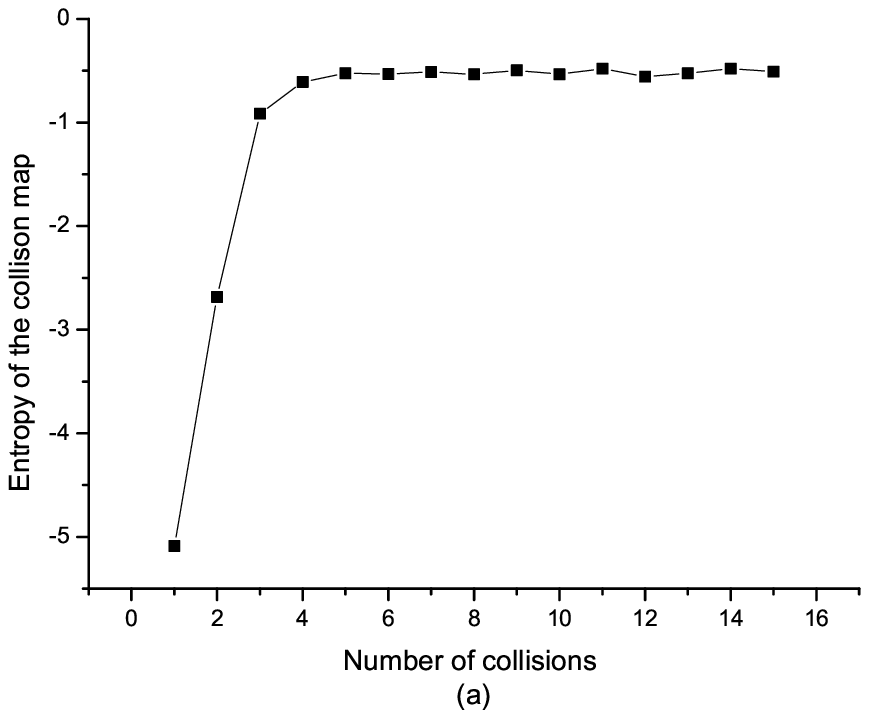}
    \includegraphics[width=65mm]{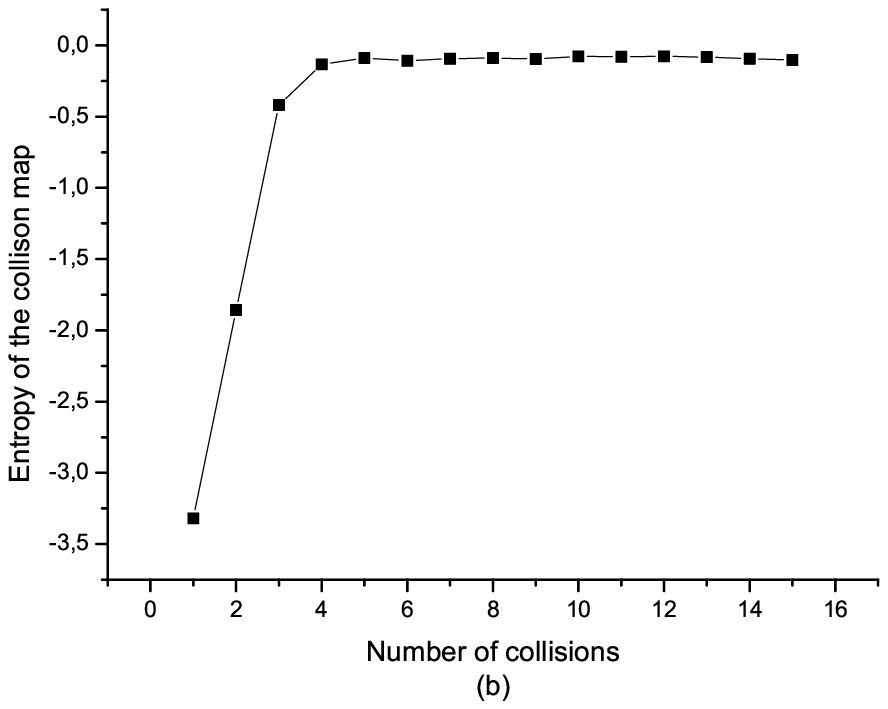}
\caption{\footnotesize{Entropy of the collision map versus number
of collisions for (a) a beam of 640 particles for a radius a=0.2,
neighboring disks centers distance 1 and a partition of
$(\beta,\psi)$ space into $25 \times 25$ cells, (b) a beam of 512
particles for the obstacles of radius 0.2, neighboring disks
centers distance 1 and a partition of $(\beta,\psi)$ space into $9
\times 9$ cells.}}\label{entropymap}
 \end{center}
\end{figure}
is the invariant measure \cite{sin1} of the cell
$\mathcal{P}_i=[\beta_i,\beta_{i+1}[ \times[\psi_i,\psi_{i+1}[$
and $\nu_t(\mathcal{P}_i)$ is the probability that a particle is
located after t collisions in $\mathcal{P}_i$ computed as
$$\frac{\mathrm{Number \,of \, particles\, in \,\mathcal{P}_i\,
having \, made \,t\,collisions} }{\mathrm{N}}.$$ The velocity
after the collision is computed from the following equation:
\begin{equation}
\textbf{V}({P}_2)=\textbf{V}({P}_1)-2(\textbf{V}({P}_1).\textbf{n})\textbf{n}
\end{equation}
where $\textbf{n}$ is the normal vector at the collision point. We
explain in the appendix the main geometric formula used for this
computation. This entropy increase is shown in the
Fig.~\ref{entropymap} for various partitions and various initial
distributions. The absolute value of the entropy of a distribution
of particles, that we call its amount of entropy, represents in fact
its distance to equilibrium. This is illustrated in the examples  of
randomly distributed initial velocity of particles having small
amount of entropy (see Fig.~\ref{entmapran} ) comparatively with
beams of particles. It is to be noted that the amount of entropy
increase under one collision is remarkably greater for the few first
ones (more or less
$2$-$4$ collisions) which corresponds to an exponential type increase (Fig.~\ref{lentropymap}).  \\
\begin{figure}[!h]
\begin{center}
  \centering
  \includegraphics[width=65mm]{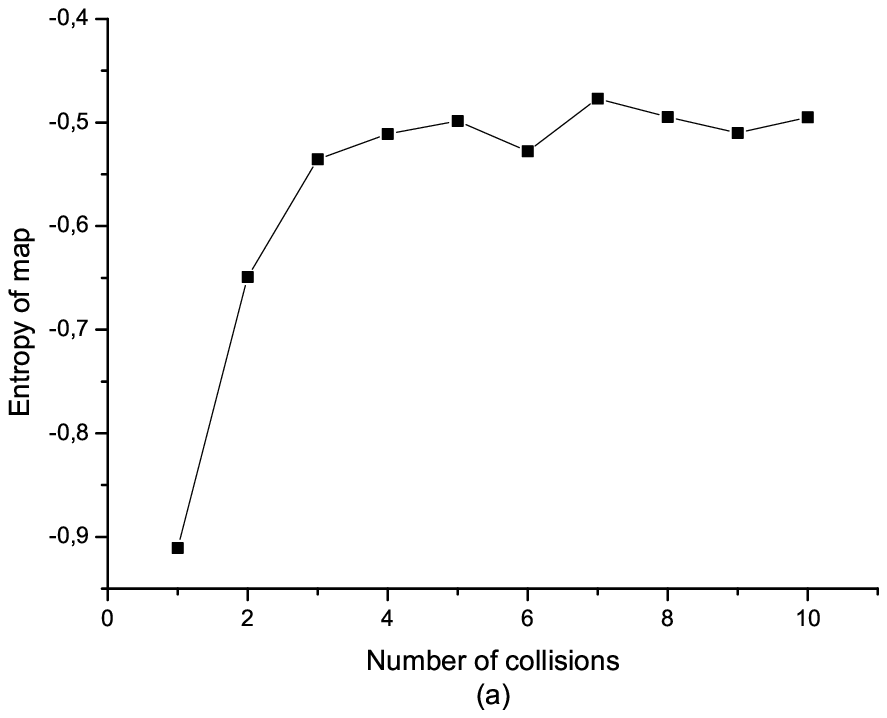}
  \includegraphics[width=65mm]{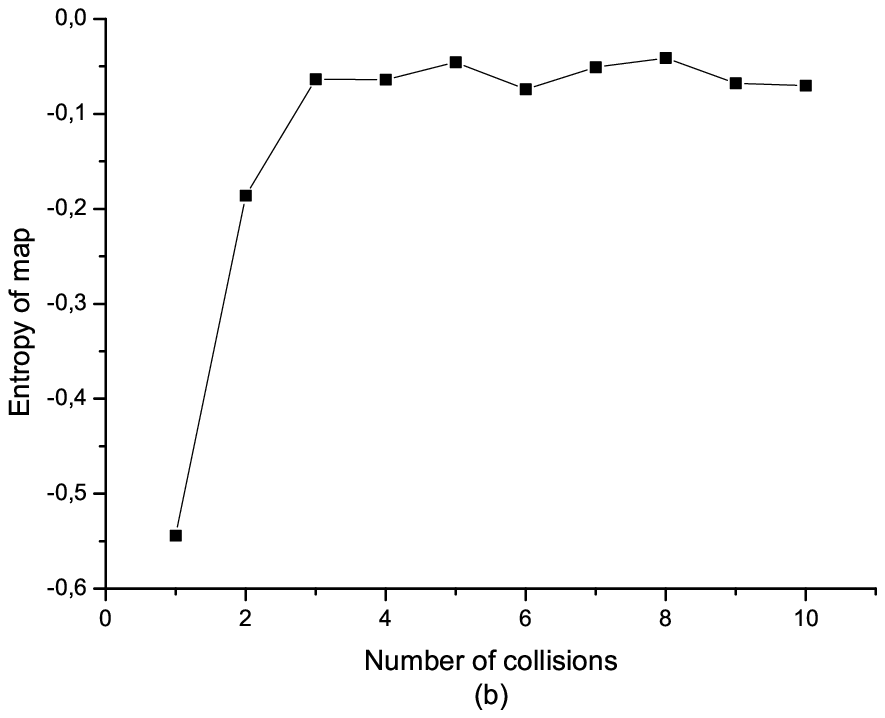}
\caption{\footnotesize{(a) and (b) are the entropy of the collision
map with random initial conditions versus number of collisions for
the system of particles of the Fig.~\ref{entropymap},
respectively.}}\label{entmapran}
 \end{center}
\end{figure}
In order to calculate Lyapounov exponents by using the method of
Benettin et al \cite{Ben},  first we calculate the Jacobian matrix
in the tangent space of the collision map:
\begin{figure}[!h]
\begin{center}
  \centering
  \includegraphics[width=65mm]{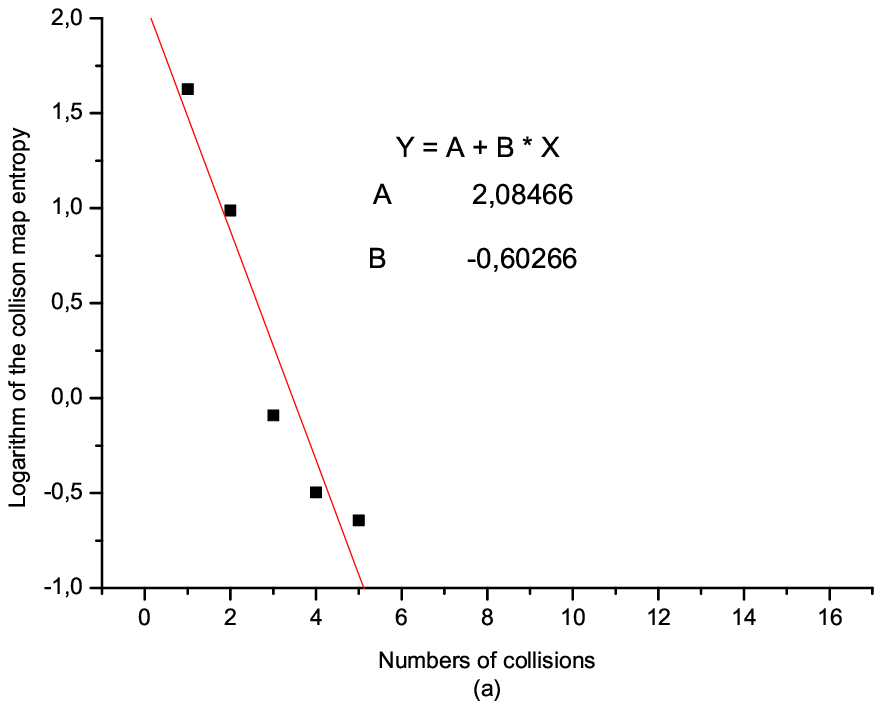}
  \includegraphics[width=65mm]{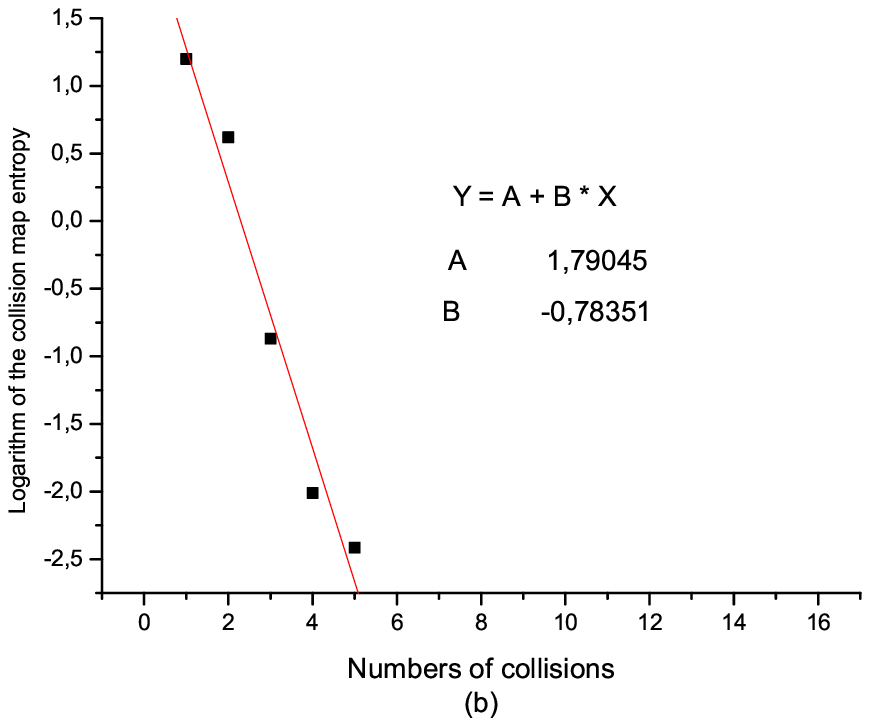}
 \caption{\footnotesize{Logarithm of the collision map entropy  versus number of
 collisions for the system of particles of the  Fig.~\ref{entropymap}.}}\label{lentropymap}
 \end{center}
\end{figure}
\[ \left(\begin{array}{cc}
\frac{\partial\beta_2}{\partial\beta_1}&\frac{\partial\beta_2}{\partial\alpha_1}\\\\
\frac{\partial\alpha_2}{\partial\beta_1}&\frac{\partial\alpha_2}{\partial\alpha_1}\\
\end{array} \right).\]
Now, comparing
$\vartriangle\mathcal{S}=\max(\mathcal{S}(t+1)-\mathcal{S}(t))$
(where the "$\max$" is taken over $t$) with the positive Lyapounov
exponent, $\lambda$, of the collision map we verify the
inequality:
\begin{equation}
 \vartriangle\mathcal{S}<\lambda
\end{equation}
as shown in Fig.~\ref{lyem}, where  this exponent is $\sim3.2$.
 The maximal entropy increase by collision  for the distribution
computed in this figure is not far from this value. So it could be
conjectured that in some suitable refinement limit, the entropy
increase of a beam tends to the positive Lyapounov exponent. The
rate of the approach to equilibrium is thus  related to the positive
Lyapounov exponent. Furthermore, the value of Lyapounov exponent is
only dependent of $\frac{D}{a}$, i.e. the ratio of the distance
between two successive obstacles over the radius of the obstacle,
and its variation is exponential as shown in Fig. \ref{lyada}.\\
\indent In order to compare the entropy increase as a function of
the collisions  with the entropy increase as a function of time,
we compute the distribution of mean free time for the first 3
collisions. From time histogram for the first three collisions of
this system ( Fig.~\ref{timeh}), we see that a great number of
particles have the same mean free time. As shown in the table
\ref{ttable}, the mean free time vary during the first three or
four collisions  but after those,  for the following collisions,
rapidly the system comes near
 the equilibrium,  where we have a constant mean free time
approximately.
\begin{table}[!h]
\begin{tabular}{|r|l|l|l|l|l|l|l|}
\hline
 Collision number&1&2&3&4&5&6&7\\
\hline Mean free time&1.966&26.174&5.801&3.820&3.611&4.452&
4.177\\
\hline
Collision number &8&9&10&11&12&13&14 \\
\hline Mean free time&4.162&4.208&4.212&3.863&4.272&4.051&4.397\\
\hline
\end{tabular}
\caption{\footnotesize{Mean free time obtained for a beam of 640
particles for a radius a=0.2, neighboring disks centers distance 1
and a partition of $(\beta,\psi)$ space $25 \times 25$
cells.}}\label{ttable}
\end{table}
\section{Spatially extended Lorentz gas entropy }
\begin{figure}[!h]
\begin{center}
  \centering
  \includegraphics[width=65mm]{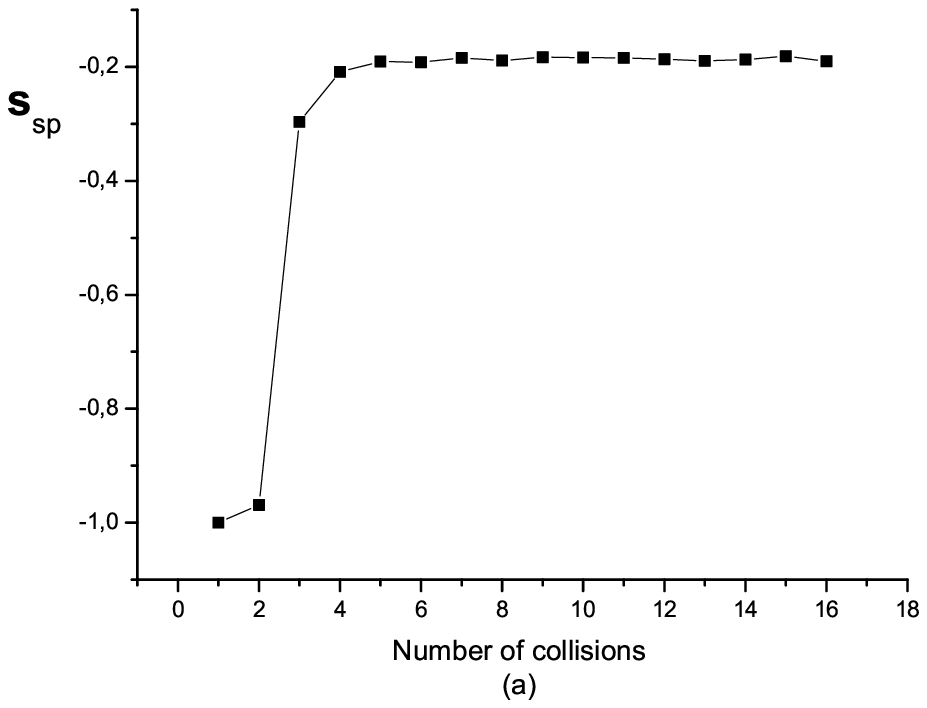}
  \includegraphics[width=65mm]{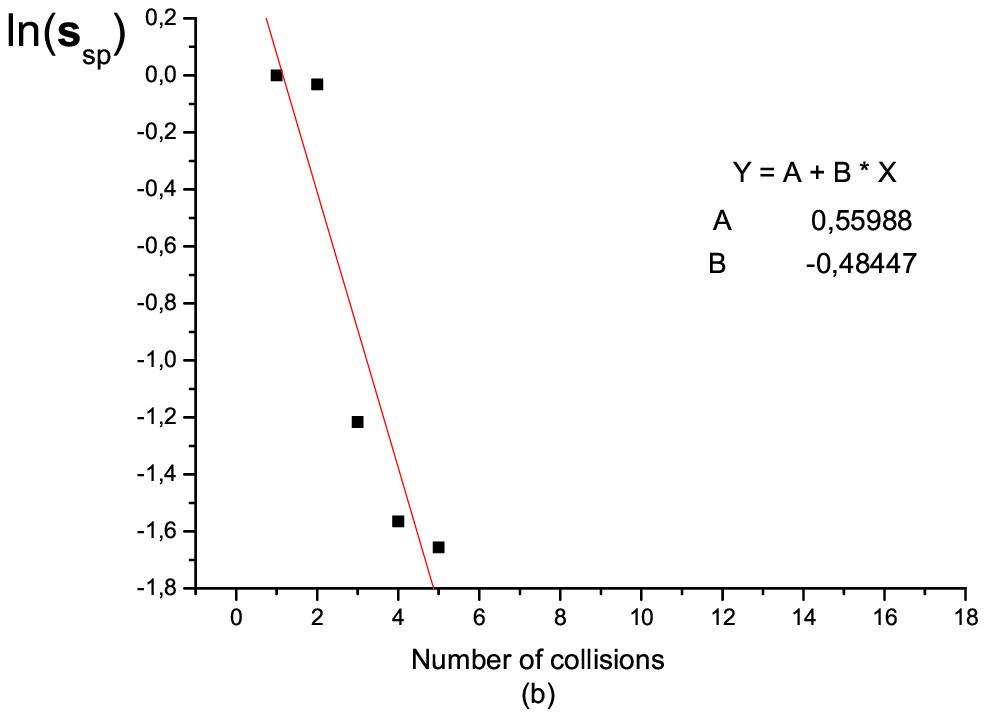}
\caption{\footnotesize{(a) Normalized space entropy of the Lorentz
gas versus number of collisions for a beam of 640 particles for
obstacles of radius a=0.2, neighboring disks centers distance 1
and a partition of $(x,y)$ space into $25 \times 25$ cells, (b)
Logarithm of the space entropy versus number of collisions for
this system.}}\label{les}
 \end{center}
\end{figure}

\indent The computation  of the  normalized space entropy equation
by using (\ref{eq2-1}) versus the number of collisions shows a
remarkable exponential increase both for beams and for random
initial distributions (Fig.~\ref{les}). The computation of sum of
the two positive Lyapounov exponents of the flow of one particle is
equal to $1.046$. Thus, we observe that the inequality between the
normalized increase of the density of the space entropy and this sum
is verified.
\section{Hard disks }\label{sechs}
\indent

\begin{figure}[!b]
\begin{center}
  \centering
  \includegraphics[width=65mm]{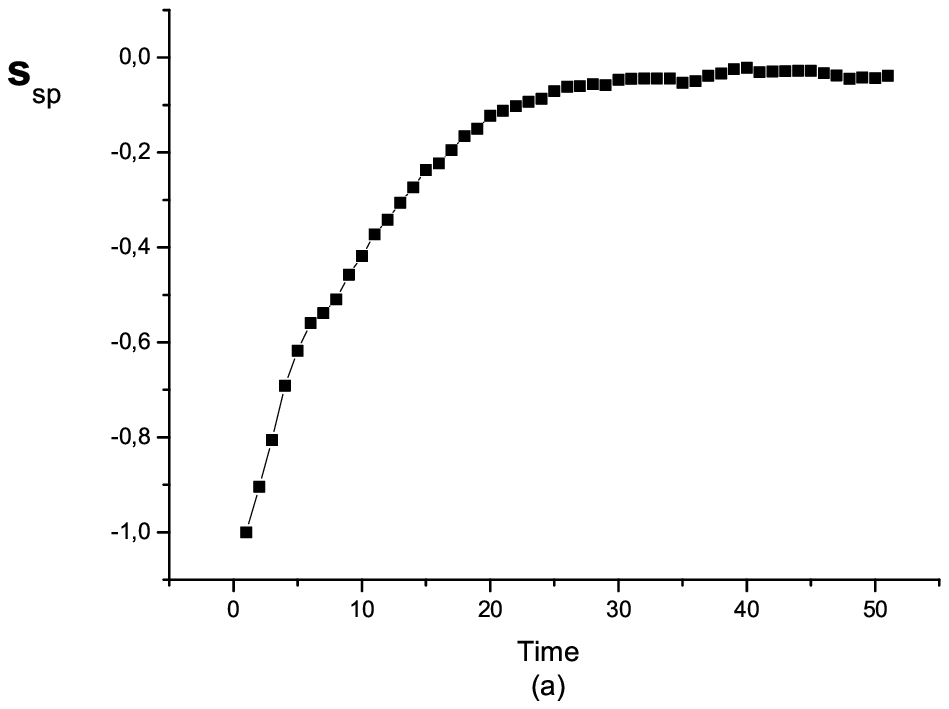}
    \includegraphics[width=65mm]{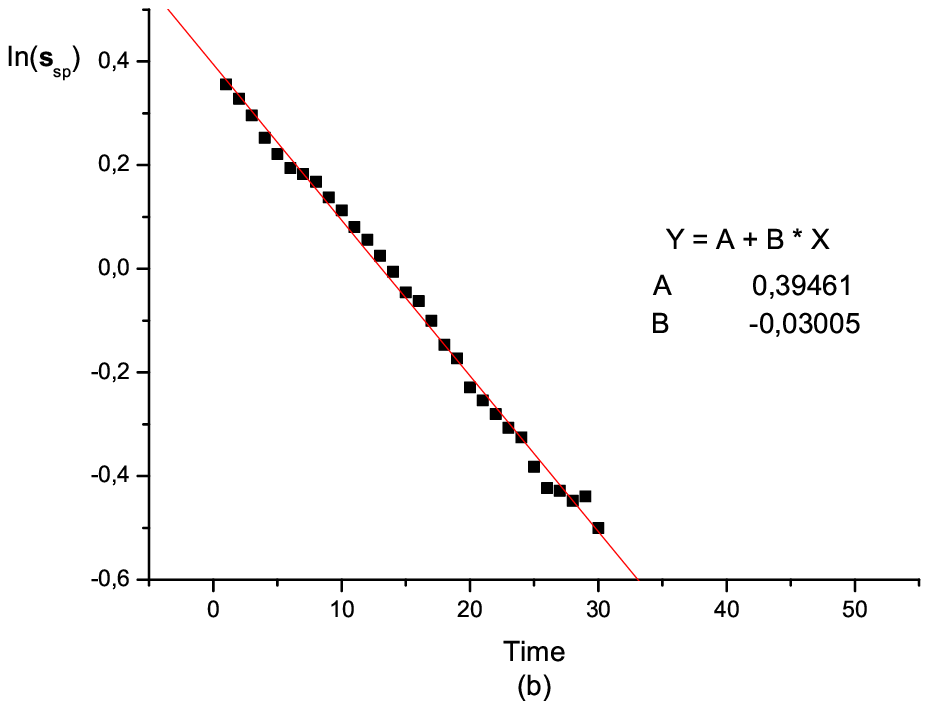}
\caption{\footnotesize{(a) Normalized space entropy and its
monotonic part logarithm of the hard disks  versus time for the
128 particles for the obstacles of radius a=0.05 which are
initially localized in the first cell of $(x,y)$ space with $6
\times 6$ cells and a density $\sigma_1=0.889$ disks per unit
area.}}\label{hes}
 \end{center}
\end{figure}

\begin{figure}[!h]
\begin{center}
  \centering
  \includegraphics[width=65mm]{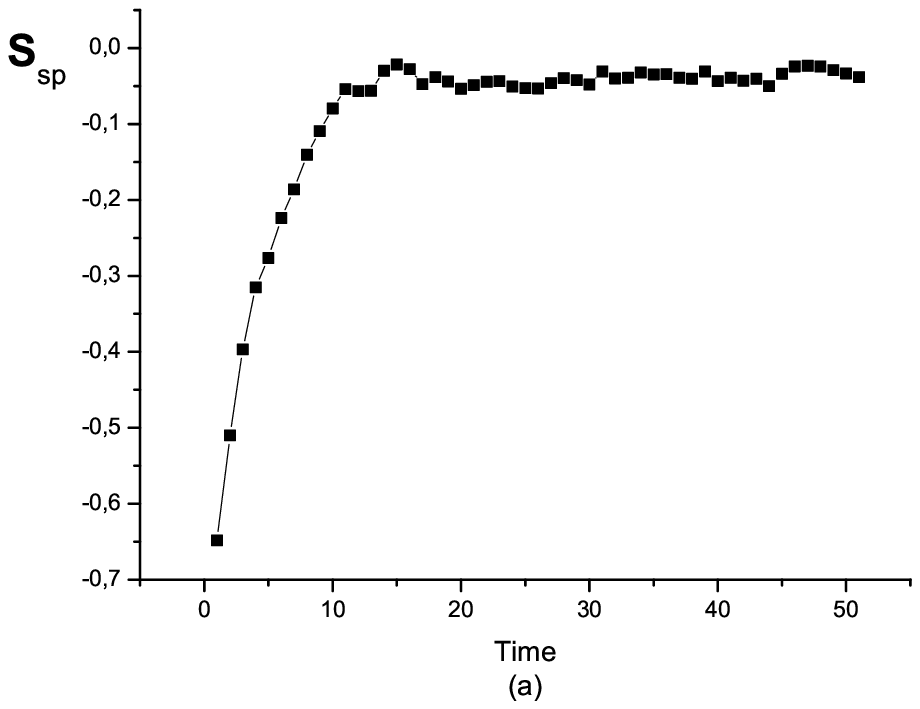}
    \includegraphics[width=65mm]{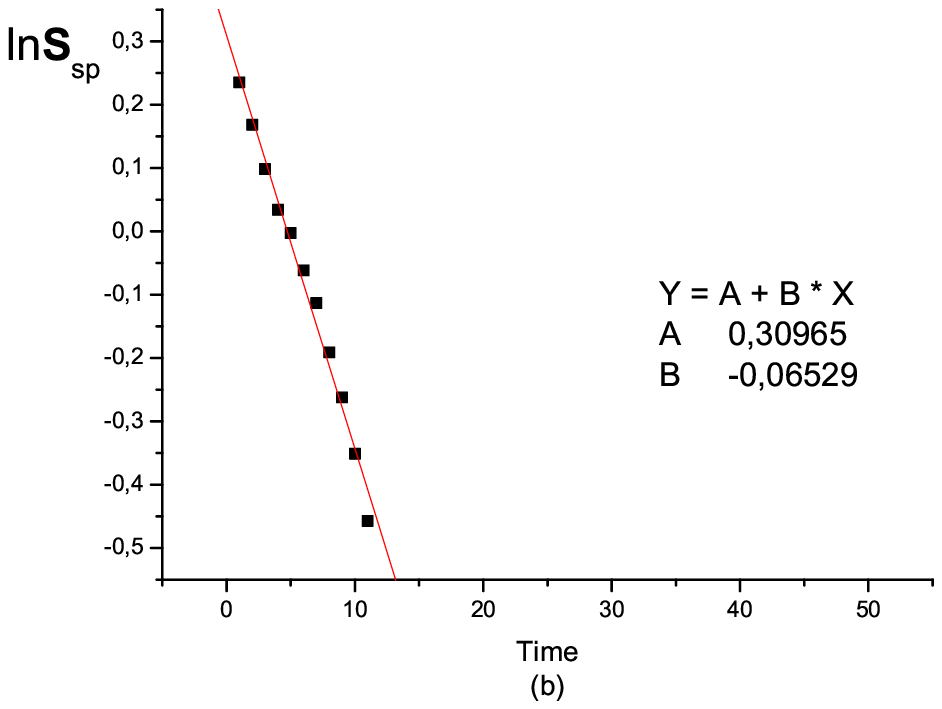}
\caption{\footnotesize{Normalized space entropy and its and its
monotonic part logarithm for the same system as Fig.~\ref{hes},
with a density $\sigma_2=3.555$ disks per unit area.}
}\label{hesh}
 \end{center}
\end{figure}

\begin{figure}[!h]
\begin{center}
  \centering
  \includegraphics[width=65mm]{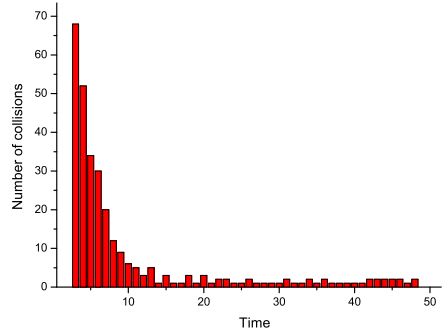}
\caption{\footnotesize{Number of collisions histogram system
versus time in Fig.~\ref{hes}.}}\label{hnc}
 \end{center}
\end{figure}

Considering a uniform space partition of a large toric space we
compute the particles densities, $\rho_i$, and the normalized
space entropy as a function of time by using the equation
(\ref{eq2-1}). Starting with a distribution of disks with
localized positions in some cell and random velocities, we compute
binary collisions instants and the trajectories of the hard disks.
These instants are determined by checking the distance between
particles, after a time interval is passed. The Lyapounov
exponents of the flow are calculated by using the Benettin et al.
algorithm. The result is shown in the Figs.~\ref{hes} and
~\ref{hesh}. These figures show the entropy  and logarithm of
monotonic part of entropy versus time of the same gas with two
distinct densities. The system in the Fig.~\ref{hesh} is more
dense than the system in Fig.\ref{hes}, and its entropy
increases more rapidly. Fig.~\ref{hnc} is a histogram of the
number of collisions, so we see that the number of collisions in a
fixed time interval is reduced for large time. From Figs.
\ref{hes} and \ref{hesh} we see that the monotonic part of the
non-equilibrium entropy is also varying exponentially with respect
to time. This shows that the collision is the main ingredient
responsible of the entropy increase as
described in the Boltzmann equation theory.\\
\begin{figure}[!h]
\begin{center}
 \centering
    \includegraphics[width=65mm]{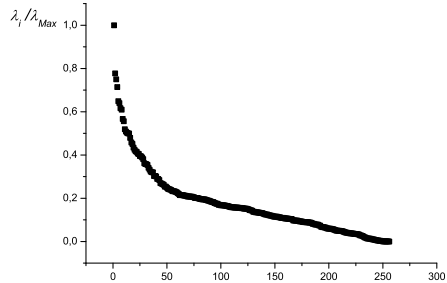}
\caption{\footnotesize{Normalized spectrum Lyapounov of exponent
of system in Fig.~\ref{hes}.}}\label{lyes}
 \end{center}
 \end{figure}
\begin{table}[!h]
\begin{center}
  \centering
\begin{tabular}{|r|l|l|}
\hline
Density&$\frac{1}{N}\sum_{\lambda_i>0}(\frac{\lambda_i}{\lambda_{max}})$&
$\triangle\textbf{s}_{sp}$\\
\hline 3.555&0.367&0.139\\
\hline 0.889&0.294&0.115\\
\hline 0.222&0.239&0.144\\
\hline
\end{tabular}
\caption{\footnotesize{The data for the hard disks systems of
radius, $a=0.05$ and the same initial conditions, with cells
$6\times6$, in terms of the density.}}\label{tably}
\end{center}
\end{table}
\indent  We shall now vary the density $\sigma=\frac{N}{V}$ and
compute the characteristic quantities. The graph of the normalized
positive Lyapounov exponents spectrum per particle for the same
system as in Fig. \ref{hes} is shown in Fig.\ref{lyes}. The
computation of the normalized sums of the positive Lyapounov
exponent,
$\frac{1}{N}\sum_{\lambda_i>0}(\frac{\lambda_i}{\lambda_{max}})$,
shows that the inequality between maximum entropy increase and the
sum of normalized  of positive Lyapounov exponents is verified (
Table \ref{tably} ).

\newpage
\section{Concluding remarks}
\indent

The computations of the entropy amount of some given
nonequilibrium initial distributions relatively to the equilibrium
measure show an exponential type increase for all considered
partitions and distributions during initial stage after which the
entropy increases slowly and fluctuates near its maximal value.
These computations confirm the existence of a relaxation time
generally assumed in the derivation of kinetic equations
\cite{balescu} and the origin of the rapid increase of the entropy
due to the number of collisions. The dispersive nature of the
obstacles is responsible of  the exponential mixing type increase.
This exponential type increase has been demonstrated for the
Sina\"{i} entropy functional \cite{sin} in hyperbolic
automorphisms of the torus. On the other hand, the relation of the
entropy increase to Lyapounov exponents can be understood through
Pesin relation and Ruelle inequality. In fact, the rate of entropy
increase should be bounded by the Kolmogorov-Sina\"{i} entropy and
such bound have been found by Goldstein and Penrose for
measure-theoretical dynamical systems under some assumptions
\cite{gold}. An open question is to characterize the measures
reaching the upper bound.

Any entropy functional  is not a completely monotonic function of time for
any dynamical system. In order to define a completely  monotonic entropy
functional for a dynamical system some conditions on the dynamics should be
imposed.  We can first suppose the map $T$ on a phase space $X$ to
be a Bernoulli system or, slightly more generally, a $K$-system.
This means that there is an invariant measure $\mu$ and some
partition $\xi_0$ of $X$ such that $T\xi_0$ becomes finer than
$\xi_0$ ( we denote it: $T\xi_0 \geq \xi_0$). Using the notation:
$T^{n}\xi_0 = \xi_n $, we obtain a family of increasingly refined
partitions, in the sense of the above order of the partitions.
Moreover, $\xi_n$ tends, as $n\rightarrow\infty$, to the finest
partition of $X$ into points, and $\xi_n$ tends, as
$n\rightarrow-\infty$, to the most coarse partition, into one set of
measure $1$ and another set of measure zero. A physical prototype of
a Bernoulli and a $K$-system is the above billiard \cite{sin1,
galla}. A geometric prototype of a Bernoulli and a $K$-system is
uniformly hyperbolic system with Sina\"{i} invariant measure
\cite{sin2}. A
non-equilibrium entropy for a family of initial measures, using
conditional expectations  relatively to the $\xi_n$ partitions was first
obtained as an equivalence between the unitary group evolution and a
semi-group of contraction operators in the space of square integrable
functions $L^2(\mu)$ successively for the baker transformation
\cite{MPC}, for Bernoulli systems \cite{CM} and for K-systems \cite{GMC}.
Its extension to the space of measures in K-systems has been realized in
\cite{C}.
 In differentiable hyperbolic dynamical systems where the fibers of the
$\xi_n$ partitions are pieces of contracting fibers, the
construction of such entropy functional results from  a
generalized coarse-graining with respect to these contracting
fibers, each fiber being a piece of manifold of zero measure.\\

\section{Appendix}
\subsection{Collision Map}
\begin{figure}[!h]
\begin{center}
  \centering
  \includegraphics[width=100mm]{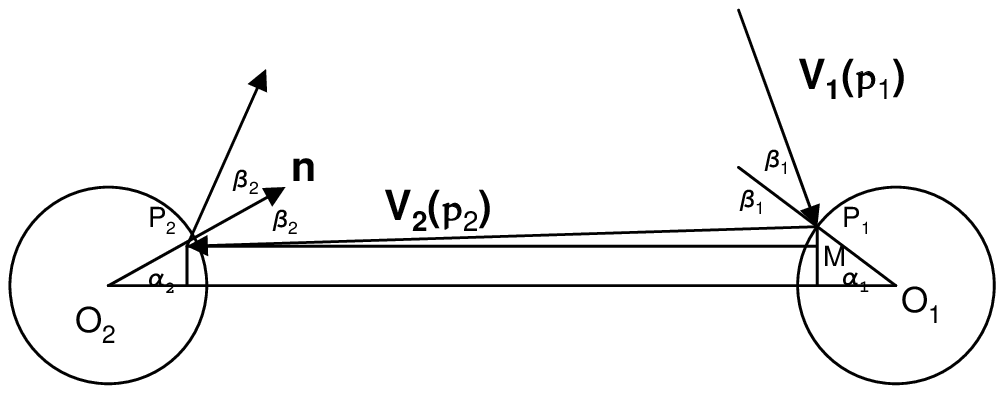}\\
 \caption{\footnotesize{non-crossing Collision.}}\label{collision1}
  \includegraphics[width=100mm]{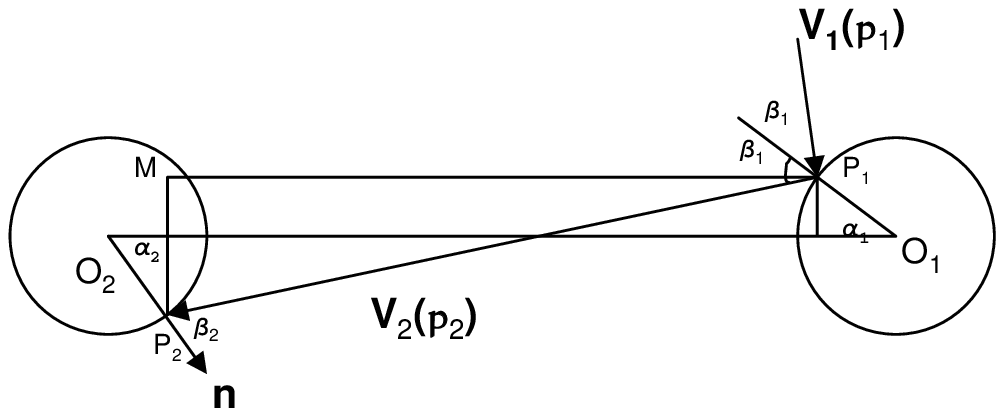}
 \caption{\footnotesize{crossing Collision.}}\label{collisionc}
 \end{center}
\end{figure}

\begin{figure}[!h]
\begin{center}
  \centering
  \includegraphics[width=100mm]{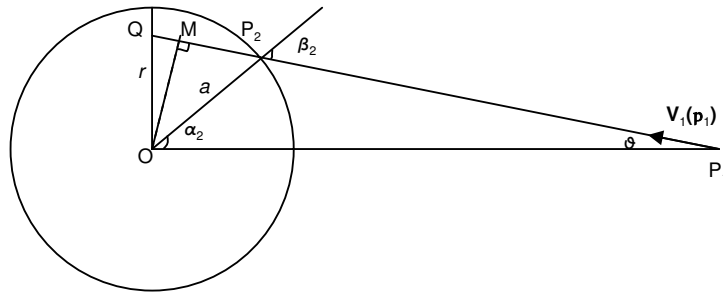}
 \caption{\footnotesize{Particle obstacle Collision.}}\label{collisionPO1}
 \end{center}
\end{figure}

\indent
 We shall give  the formula of the \emph{collision
map}. We consider a particle which undergoes the first collision
with the disk of center $\mathrm{O}_1$ with velocity
$\textbf{V}_{1}(p_1)$ and the second collision with the disk of
center $\mathrm{O}_2$ with velocity $\textbf{V}_{1}(p_2)$. Two
cases are possible. First, we consider non-crossing of the centers
line as in the Fig.~\ref{collision1}. In this figure the angle
$\widehat{\mathrm{P_1P_2}M}$ is
$\alpha_2-\beta_2=-(\alpha_1-\beta_1)$, where $\mathrm{M}$ is such
that $\mathrm{MP_2}$ is parallel to $\mathrm{O_1O_2}$. We can
write
\begin{equation}
\mathrm{P_1M}=\mathrm{P_1P_2}\cos(\beta_1-\alpha_1)=d-a\cos\alpha_1-a\cos\alpha_2.
\end{equation}
and
\begin{equation}
\mathrm{P_2M}=\mathrm{P_1P_2}\sin(\beta_1-\alpha_1)=a\sin\alpha_1-a\sin\alpha_2,
\end{equation}
if we eliminate $\alpha_2$ between these equations we arrive at
\begin{equation}
\beta_2=\arcsin[\frac{d}{a}\sin(\beta_1-\alpha_1)+\sin\beta_1].\label{eq3}
\end{equation}
In crossing case which we present in Fig.~\ref{collisionc} we see
that the angle $\widehat{\mathrm{P_2P_1M}}$ is equal to
$\alpha_2-\beta_2=\alpha_1-\beta_1$, and the length of
$\mathrm{P_2M}$ is changed to:
\begin{equation}
\mathrm{P_2M}=\mathrm{P_1P_2}\sin(\beta_1-\alpha_1)=a\sin\alpha_1+a\sin\alpha_2,
\end{equation}
then, we have
\begin{equation}
\beta_2=\arcsin[\frac{d}{a}\sin(\beta_1-\alpha_1)-\sin\beta_1].\label{eq4}
\end{equation}
To obtain $\beta_2$ in the first collision between particle and
obstacle Fig.~\ref{collisionPO1} , we take $d=\mathrm{OP_1}$,
$\beta_1=0$ and $\alpha_1=\vartheta$ in the collision map.
\subsection{Algorithm description}
\begin{figure}[!h]
\begin{center}
  \centering
  \includegraphics[width=100mm]{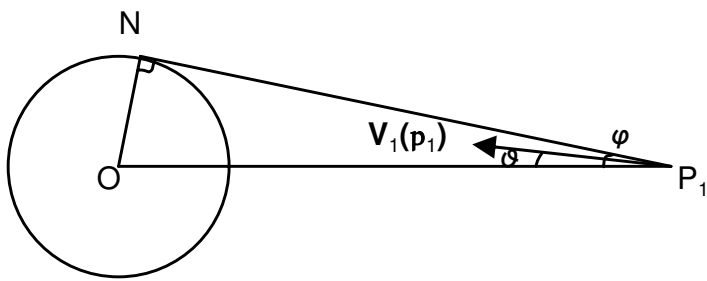}
 \caption{\footnotesize{Particle obstacle Collision.}}\label{collisionPO}
 \end{center}
\end{figure}

\begin{figure}[!h]
\begin{center}
  \centering
  \includegraphics[width=100mm]{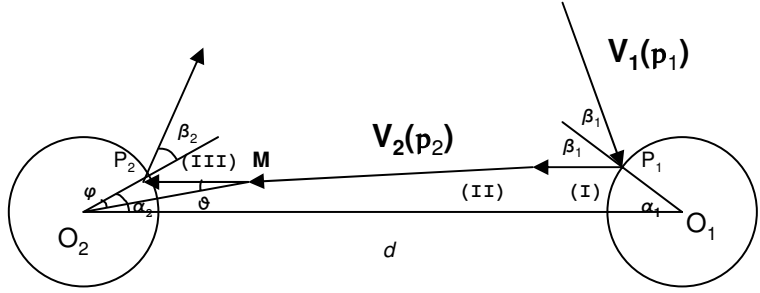}
 \caption{\footnotesize{The motion of the particle take place on a tours.}}\label{collisionb}
 \end{center}
\end{figure}

In this section we describe the algorithm which we used in our
program for Lorentz gas. We first define in the \emph{main} of our
program the initial conditions for the particles and the obstacles
positions. In the second step, we compute with which obstacle, a
particle will collide: we measure the angle between velocity of
particle and the line between this particle and the center of
obstacle, $\mathrm{OP}_1$ in Fig.~\ref{collisionPO}, if this angle
is less than or equal to the angle between this line,
$\mathrm{OP}_1$, and the tangent line on the circle, i.e.
$\mathrm{P_1N}$, in brief if $\vartheta \leq \varphi$ in
Fig.~\ref{collisionPO}, we have a collision. Now, we use the
collision map equation (\ref{eq3}) or (\ref{eq4}) to obtain the
collision angles, $\beta_2$, and $\alpha_2$ (see
Fig.~\ref{collisionPO1}). In this step, we can also obtain the
length of arrow of our induced collision map, i.e. $\mathrm{P_1P_2}$
(see Fig.~\ref{collisionPO1}), easily as:
\begin{equation}
\mathrm{P_1P_2}=\frac{\mathrm{OP_1}- a
\cos\alpha_2}{\cos\vartheta}
\end{equation}
where $\alpha_2=\beta_2-\vartheta$. Then, we can calculate the time
of flight of particle  between two collisions, respectively, as
$t=\mathrm{P}_1\mathrm{P}_2/v$. This provides the trajectory of a
particles.\\
Let us turn  the computation of space entropy. When a particle
arrives at a wall of the big torus, before it does a collision with
an obstacle (see on the Fig.~\ref{collisionfb}) trajectories are
pursued until it undergoes a collision on the torus. We have to
compute  the position of the obstacle that the particle will hit
(see Fig.~\ref{collisionb}) and the angle $\alpha_1$ in the
collision map, and to determine which type of collision, i.e.
crossing or non-crossing case, will occur. We first find the angle
of collision
\begin{equation}
\beta_2=\vartheta+\varphi=\arcsin[\frac{\mathrm{MO}_2}{a}\sin\vartheta],
\end{equation}
then we arrive at $\alpha_1$ and $\alpha_1$ as
\[ \left \{\begin{array}{ll}
\alpha_1^n=\beta_1+\arcsin[\frac{a}{d}(\sin\beta_2-\sin\beta_1)],&
\alpha_2^n=\beta_2+(\beta_1-\alpha_1)\\
\alpha_1^c=\beta_1-\arcsin[\frac{a}{d}(\sin\beta_2+\sin\beta_1)],&
\alpha_2^c=\beta_2-(\beta_1-\alpha_1)\\
\end{array} \right. \]
\begin{equation}
\end{equation}
where the superscript "$c$" corresponds to crossing case and "$n$"
corresponds to non-crossing case see equations (\ref{eq3}) and
(\ref{eq4}), respectively. In the above equations, the $d$
parameter is unknown, and will be recognized it in the end of this
appendix. If we subtract the above equations we obtain
\begin{equation}
\alpha_1^c-\alpha_1^n=-\arcsin[\frac{a}{d}(\sin\beta_2+\sin\beta_1)]-\
arcsin[\frac{a}{d}(\sin\beta_2-\sin\beta_1)]
\end{equation}
We can see the above equation yields $\alpha_1^c-\alpha_1^n\leq0$.
It means that in the same conditions the angle $\alpha_1$ in the
non-crossing is greater than crossing case. Also, we can get the
same conclusion for $\alpha_2$, i.e. $\alpha_2^c\leq\alpha_2^n$.
Now, we initiate the algorithm in the non-crossing case and we
find $\alpha_1^c$ and $\alpha_2^n$. If $\vartheta\leq\alpha_2$,
thus, we had a correct supposition, otherwise, we must consider
the crossing case, and we re-calculate these angles. In order to
find in this case the parameter $d=|\mathrm{O}_1\mathrm{O}_2|$, we
calculate it by approximation method. The equation that recognize
$d$ is:
\begin{equation}
d=vdt\cos(\beta_1-\alpha_1)+a(\cos\alpha_1+\cos\alpha_2)
\end{equation}
where $dt$ is the time of free flight of particle between two
collisions, see Figs. (\ref{collisionfb} and \ref{collisionb} ).
In the above equation we have two unknown variables, $\alpha_1$
and $\alpha_2$. We use the \emph{zeroth approximation} as
\begin{equation}
d\approx vdt
\end{equation}
where we used $a\ll vdt$. Now, we calculate the angles, $\alpha_1$
and $\alpha_2$,  as mentioned in above of this appendix. Then, we
re-calculate $d$ with the \emph{first approximation}, and we can
repeat this procedure. However, the convergence is very rapid.

\end{document}